\documentclass[prl,showpacs,twocolumn,letterpaper]{revtex4}
\usepackage{graphics}
\newcommand{\be}{\begin{equation}}
\newcommand{\ee}{\end{equation}}

\def\br{{\bf r}}

\def\bea{\begin{eqnarray}}
\def\eea{\end{eqnarray}}
\def\ben{\begin{equation}}
\def\een{\end{equation}}
\def\benu{\begin{enumerate}}
\def\enu{\end{enumerate}}

\def\sss{\scriptscriptstyle\rm}

\def\1var{(\bx_1...\bx\N)}

\def\half{\frac{1}{2}}

\def\br{{\bf r}}

\def\bx{{\br t}}

\def\xc{_{\sss XC}}

\def\N{_{\sss N}}

\def\SPA{^{\rm SPA}}

\def\KS{^{\rm KS}}

\def\sph_int{ {\int d^3 r}}

\begin{document}
%\sf%\large

\title{Double-Pole Approximation in
Time-Dependent Density Functional Theory}
\author{H. Appel}
\author{E.K.U. Gross}
\affiliation{Fachbereich Physik, Freie Universit\"{a}t Berlin, Arnimallee 14, D-14195 Berlin-Dahlem, Germany.}
\author{K. Burke}
\affiliation{Department of Chemistry and Chemical Biology, Rutgers University, 610 Taylor Rd, Piscataway, NJ 08854}
\date{\today}
\begin{abstract} 
%\sf%\large
A simple approximate solution to the linear response equations
of time-dependent density functional theory (TDDFT) is given.  This extends
the single-pole approximation (SPA) to two strongly-coupled poles.
The analysis provides both an illustration of how TDDFT  works when
strong exchange-correlation effects are present and insight into such
corrections.  
For example, interaction can cause a transition to vanish entirely
from the optical spectrum.
\end{abstract}

\maketitle

%%%%%%%%%%%%%%%%%%%%%%%%%%%%%%%%%%%%%%
%%%%%%%%%%% INTRODUCTION %%%%%%%%%%%%%
%%%%%%%%%%%%%%%%%%%%%%%%%%%%%%%%%%%%%%
\section{introduction}
\label{s:intro}

Ground-state density functional theory (DFT) 
has been very successful
for atoms, molecules, and solids \cite{KS65,FNM03}. 
Similar success is now being enjoyed by 
time-dependent DFT (TDDFT) \cite{RG84,BWG05}, because of its combination
of accuracy combined with low computational cost \cite{FR05}.
While TDDFT has a huge variety of applications \cite{MG04}, it is low-lying
photo-excitations of molecules that has seen its greatest use \cite{FR05}.

In the present work, we restrict our discussion to linear response
of a non-degenerate ground-state.
Just as in ground-state DFT, all many-body effects, i.e., exchange and correlation
(XC), are contained in a well-defined functional, the XC kernel \cite{GK85}.
In any practical calculation, this functional must be approximated.  
In most calculations, an adiabatic
approximation is made, and the static limit of the kernel is applied.
Typical approximations are then adiabatic local density approximation \cite{GK85}
(ALDA) or generalized gradient approximation, or exact exchange \cite{UGG95,PGG98,G98}.
The reliability and accuracy of these approximations to TDDFT is much
less well-understood than it is in ground-state DFT.

One can do many calculations on many systems, in order to gain insight
into the accuracy and reliability of theory,
but it can be much more effective to develop simple
approximations to the solution of the TDDFT response problem \cite{AGB03}.
A classic example is the single-pole approximation \cite{PGG96}, within
which TDDFT yields a simple correction to the KS transition frequencies
which is just the expectation value of the Hartree-XC kernel on the transition
orbitals.   While usually accurate \cite{AGB03}, the most important
feature of this approximation is the insight it yields into the
workings of TDDFT.   It yields a first approximation to TDDFT effects
with almost no extra effort beyond a ground-state calculation, and 
gives a simple picture for such effects \cite{WMB05}.  It has also been shown \cite{AGB03}
that, if a transition is only weakly-coupled to others in the system,
one can use this to estimate the XC kernel itself.  Unfortunately,
this is rarely the case in practice.

In the present work, we generalize the SPA to a {\em double-pole} approximation
(DPA), in which we explicitly solve the TDDFT response equations for
exactly two transitions.  This produces a variety of results beyond that
of SPA.  Most importantly, one can study TDDFT XC corrections to
KS levels when there is strong coupling between levels.  But one can also
see when SPA fails, and recover G\"orling-Levy perturbation theory \cite{GL93} results
for the coupling-constant expansion of excited
states \cite{AGB03}.  DPA has recently
been successfully applied to  core-hole interaction in the X-ray
absorption spectroscopy of 3d transition 
metals \cite{SGAS05}.

\section{Double-pole approximation}
\label{s:dpa}

In the matrix-formulation of the TDDFT response equation within the adiabatic
approximation, the exact eigenvalues and oscillator
strengths can be obtained from the solution of the following eigenvalue problem \cite{C96}
\be
\label{Casida}
\sum_{q'} {\tilde W}_{qq'}(\Omega) \ v_{q'} = \Omega^2\ v_{q},
\ee
where the matrix $W$ is given by
\be
{\tilde W}_{qq'}(\Omega)
= \omega^2_q \, \delta_{qq'} + 4\,{\sqrt{\omega_q\, \omega_{q'}}}\,
M_{qq'}(\Omega)
\label{CasidaMatrix}
\ee
and
\be
M_{qq'}(\Omega)
= \int d^3r\int d^3r'\, \Phi^*_q(\br)\,
K(\br\br'\Omega)\,\Phi_{q'}(\br').
\label{Mdef}
\ee
Here $\omega_i$ is the Kohn-Sham transition frequency and for
single particle transitions $q$ ($q\equiv k\to j$) the shorthand
$\Phi_q(\mathbf{r}):=\varphi_k(\mathbf{r})\varphi_j^\ast(\mathbf{r})$
has been introduced. The kernel $K(\mathbf{r},\mathbf{r}^{\;\prime},\omega)$ consists 
of the bare Coulomb interaction and the approximate XC kernel  
$f\xc(\mathbf{r},\mathbf{r}^{\prime}, \omega)$:
\be
K(\mathbf{r},\mathbf{r}^{\prime},\omega)=
\frac{1}{\left|\mathbf{r}-\mathbf{r}^{\prime}\right|}+
f\xc(\mathbf{r},\mathbf{r}^{\prime},\omega). 
\label{Kernel}
\ee
Atomic units ($e^2=\hbar=m=1$) are used throughout.

We now solve these equations exactly for a $2\times2$-system, i.e., ignoring
coupling to all other transitions.  
% Note, we use +/- and j notation at the same time. Also the figure with the levels
% is gone ...
%We label transitions and oscillator strengths with 1 and 2 (see Fig. 1).
To simplify the discussion we assume a frequency independent kernel and
real orbitals, i.e. $M_{qq'}=M_{q'q}$.
Thus the relation between matrix elements of Casida's equation and
the kernel is:
\ben
W_{ii}=\omega_i^2 + 4\, \omega_i\; M_{ii},~~~~~W_{12}=4\,{\sqrt{\omega_1\omega_2}}\, M_{12}.
\label{Wdef}
\een
Next define the average
\ben
\label{barWdef}
{\overline W} = \half\,(W_{11}+W_{22}) 
\een
and difference
\ben
\label{DWdef}
\Delta W = W_{22}-W_{11}
\een
of the diagonal elements.
We define a mixing angle by:
\ben
\label{TanTheta}
\tan\theta = \frac{2\, W_{12}}{\Delta W},
\ee
choosing the branch between $0$ and $\pi$.
The eigenvalues can then be written succinctly as
\be
\label{Eigenvalues}
\Omega_\pm^2 =  \overline W \pm \half\,\frac{\Delta W}{\cos\theta},
\een
while the {\em normalized} eigenvectors are
\renewcommand{\arraystretch}{1.5}
\be
\vec v_+ = \left( \begin{array}{c} \sin\frac{\theta}{2} \\ \cos\frac{\theta}{2} \end{array} \right), \qquad
\vec v_- = \left( \begin{array}{c} -\cos\frac{\theta}{2} \\ \sin\frac{\theta}{2} \end{array} \right). \qquad
\ee
The physical oscillator strength can be obtained from the following expression  \cite{C96}
\be
\label{osc}
f_\pm = \frac{2}{3}\, | \vec x^T\,S^{-\half}\, \vec v_\pm |^2,
\ee
where
\be
S^{-\half} = \left(
\begin{array}{cc} 
  \sqrt{\omega_1} & 0 \\
  0               & \sqrt{\omega_2} 
\end{array}
\right), 
\qquad \vec x = \left( \begin{array}{c} x^{KS}_1 \\ x^{KS}_2 \end{array} \right),
%              = \left( \begin{array}{c} \la 3\p x\p 2\ra \\ \la 3\p x\p 1\ra\end{array} \right)
\ee
and the $x^{KS}_j$ denote dipole matrix elements of KS orbitals. 
Given that there are only two transitions, we give 
a geometric meaning to the oscillator strengths.  Writing
\ben
f_1\KS=\sin^2\alpha\KS,~~~~f_2\KS=\cos^2\alpha\KS
\label{alphaKS}
\een 
and
\ben
f_-=\sin^2\alpha,~~~~f_+=\cos^2\alpha,
\label{alpha}
\een 
we find
\ben
\alpha=\alpha\KS - \theta/2,
\label{physOsc}
\een
i.e., the oscillator strengths are represented by a unit vector in 2d space,
and the coupling merely rotates this vector. 
Note that the Thomas-Reiche-Kuhn (TRK) sum rule (sum of the
oscillator strengths is 1) is obviously preserved.

\section{Single-pole approximation}
\label{s:spa}

As mentioned above, the single-pole approximation is a useful
approximation to TDDFT results.  We recover SPA results by
inserting $\theta=0$ in our formulas.  Thus
\be
\label{EigenSPA}
\Omega\SPA_\pm = \sqrt {\overline W \pm \frac{\Delta W}{2}}
\een
and the oscillator strengths reduce to their KS values.

We can now study the leading corrections to SPA produced by DPA
when the coupling between poles is weak.  Writing $\eta=W_{12}/\Delta W$,
and assuming $\eta << 1$, 
for the eigenvalues, we find
\be
\label{SPAplus}
\Omega\pm=\Omega\SPA_\pm \pm \frac{W_{12}}{2\, \Omega\SPA_\pm }\eta  +O(\eta^2),
\een
while for the oscillator strengths, we have
\bea
\label{physOscplus}
f_+
    & = & f_2^{KS} + 2 \eta \sqrt{f_1^{KS} f_2^{KS}} +O(\eta^2),\nonumber\\
f_- & =
    & f_1^{KS} - 2\eta  \sqrt{f_2^{KS} f_1^{KS}}+O(\eta^2).
\eea
Note that the corrections to the peak positions are
second-order in $W_{12}$, while the corrections to $\sqrt{f_\pm}$
are first-order.   Thus SPA is expected to be much better for
peak positions than for peak heights.

Lastly, we point out that this expansion was deduced for the
general case in Ref. \cite{AGB03}, and used (among other things)
to identify coefficients in the G\"orling-Levy expansion of
excited-state energies.  Our results here agree with those,
but in the special case of transitions to which DPA applies,
yield results that include a resummation of {\em all} orders in the
adiabatic  coupling
constant of density functional theory.

\section{High-frequency limit}
\label{s:hi}

So far we have given exact results for the double-pole
approximation.  However, in many cases where DPA applies, there is
a further simplification.   Usually the two transitions are 
closer to each other than any others that couple to the pair. If in 
addition their frequency difference is small relative to their mean
frequency, for both the interacting and KS systems, i.e.,
\ben
\overline \Omega,\, \overline \omega >> \Delta\Omega,\, \Delta\omega,
\een
we find much simpler results, which are
very useful for interpretation.

The SPA discussed above reduces to
\ben
\Omega\SPA_\pm=\omega_i + 2\,M_{ii},~~~~(i=1,2).
\label{OmegaSPAhi}
\een
In fact, the original SPA was applied for just a forward transition,
yielding exactly this result \cite{PGG96}.
However, this approximation violates the TRK sum rule when applied
outside the high-frequency limit,
so the symmetric result (sometimes called the small-matrix
approximation \cite{VOC99,GPG00}) is preferable.  We use the term SPA to mean the
symmetric result throughout this paper.

The mixing angle is given by
\ben
\tan\theta= \frac{4 M_{12}}{\Delta\Omega\SPA},
\label{thetahi}
\een
i.e., it is the ratio of the off-diagonal matrix elements of
the kernel on the scale of the separation in SPA that matters.
We find
\ben
\Omega_\pm = \overline{\Omega\SPA}\pm \frac{\Delta\Omega\SPA}{2\cos\theta}.
\label{Omegahi}
\een
SPA yields the correct average position of the two lines, but
their splitting is greater than SPA predicts (level repulsion).

\section{Illustrations}
\label{s:illust}

\begin{figure}[t]
  \unitlength1cm
  \begin{picture}(5.5,5.5)
    \put(-3.5,-8.5){
      \includegraphics{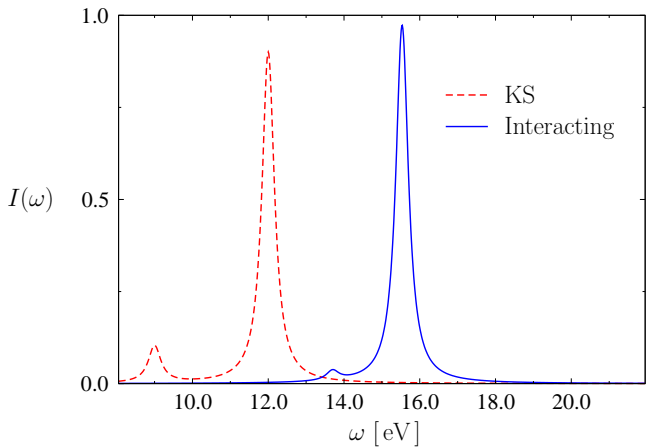}
    }
  \end{picture}
   \vspace{0.2cm}
  \caption{Interacting and Kohn-Sham spectra as function of frequency 
    ($\omega_1$ = 9 eV, $M_{12} = 0.2$ eV).}
   \label{fig5a} 
   \vspace{0.4cm}
\end{figure}
\vspace{0.3cm}
\begin{figure}[ht]
  \unitlength1cm
  \begin{picture}(5.5,5.5)
    \put(-3.5,-8.5){
      \includegraphics{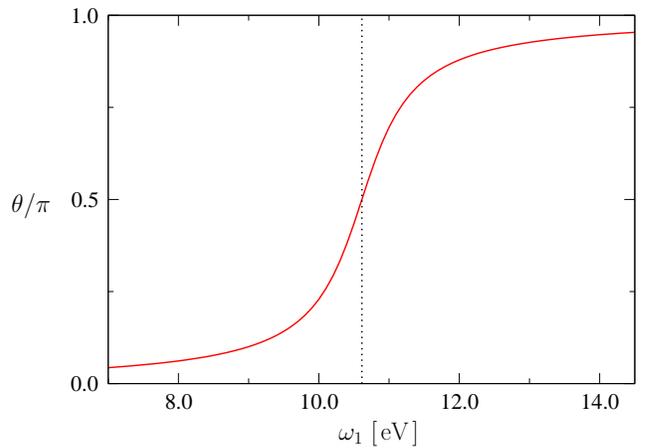}
    }
  \end{picture}
   \vspace{0.2cm}
  \caption{The scaled coupling angle $\theta/\pi$ as function of the position
of the lower transition.}
   \label{fig2} 
   \vspace{0.4cm}
\end{figure}
To illustrate our results, consider a weak lower-frequency
transition ($\omega_1 = 9$ eV, $f_1^{KS}=1/10$) and  a strong higher-frequency transition
($\omega_2=12$ eV, $f_2^{KS}=9/10$).  We imagine these have significant diagonal kernel
matrix
elements $M_{11}=3$ eV, $M_{22}=2$ eV, but are not strongly coupled to one
another, $M_{12} = 0.2 $eV. We have plotted the interacting and KS spectra in Fig. \ref{fig5a}.
The peaks are Lorentzians of width 0.2, mimicking a measurement of finite
resolution.
Because the coupling is weak, the single-pole
approximation is excellent, and accurately predicts the large shifts in
positions.  However, SPA wrongly predicts no variation in oscillator strength.
In fact, one can see from the figure that the first peak has actually
{\em lost} intensity relative to its KS value.
%%%%%%%%%%%%%%%%%%%%%%%%%%%%%%%%%%%%%%%%%%%%%%%%%%%%%%%%%%%%%%%%%%%%%%

In the rest of this section, we explore what happens in the DPA model of
TDDFT.
In order to emphasize that it is not the absolute magnitude of the
off-diagonal matrix element that is significant, but rather its strength
relative to the separation between the peaks, we now consider all the
same parameters, but imagine increasing $\omega_1$.  In Fig. \ref{fig2},
we plot the mixing angle as a function of $\omega_1$.
At $\omega_c=2\,(-3+\sqrt{69})~eV \approx 10.61~eV$, 
the diagonal matrix elements $W_{ii}$ match,  so that
$\Delta W=0$ and $\theta=\pi/2$.  At that point, the peaks are
a 50:50 mixture of the two KS levels.   
\begin{figure}[b]
  \vspace{0.4cm}
  \unitlength1cm
  \begin{picture}(5.5,5.5)
    \put(-3.5,-8.5){
      \includegraphics{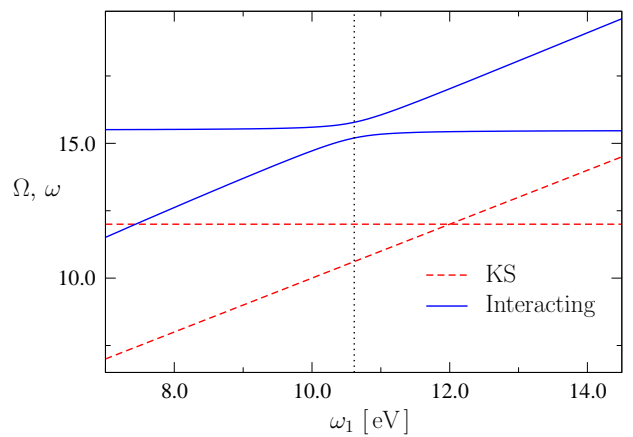}
    }
  \end{picture}
   \vspace{0.2cm}
  \caption{Interacting and Kohn-Sham excitation energies as function 
  of $\omega_1$.}
   \label{fig4a} 
\end{figure}
In that region, the levels are strongly
coupled, and the spectrum distorts mightily from its KS shape.
The width of the transition region can be defined as the change
in frequency needed to bring $\theta$ from $\pi/4$ to $3\pi/4$,
and, from Eq. (\ref{thetahi}) in the high-frequency limit, is seen to be
$4 M_{12} = 0.8$ eV here, i.e., proportional to the off-diagonal
element, but quite a bit larger.  More significantly, there
are tails in the transition that decay extremely slowly with
pole separation.
On the contrary, SPA yields a function that steps from 0 to 1 at 11 eV.

To see this,
in Fig. \ref{fig4a}, we plot
the interacting levels $\Omega_\pm$ as a function of $\omega_1$, and
observe the avoided crossing.  Note that straight line plots, extrapolated
from the limits where $\omega_1$ is either far above or far below
$\omega_c$, yield extremely accurate results almost everywhere.  This
is the SPA result.  In fact, from Eqs. (\ref{Eigenvalues}) and (\ref{EigenSPA}),
we see that the crossover point is {\em exactly} given by SPA. 
Moreover, in the high-frequency limit, Eqns. (\ref{OmegaSPAhi}-\ref{Omegahi})
yield
\ben
|\Delta\Omega|^2=|\Delta\Omega\SPA|^2 + 16\,|M_{12}|^2.
\label{DOmega}
\een
So if the off-diagonal matrix elements are small relative
to the SPA separation, the true separation is not much greater;
the closest the two levels come is a separation of 4\,$|M_{12}|$,
i.e., they never cross.

\begin{figure}[b]
  \unitlength1cm
  \begin{picture}(5.5,6.5)
    \put(-3.5,-8.0){
      \includegraphics{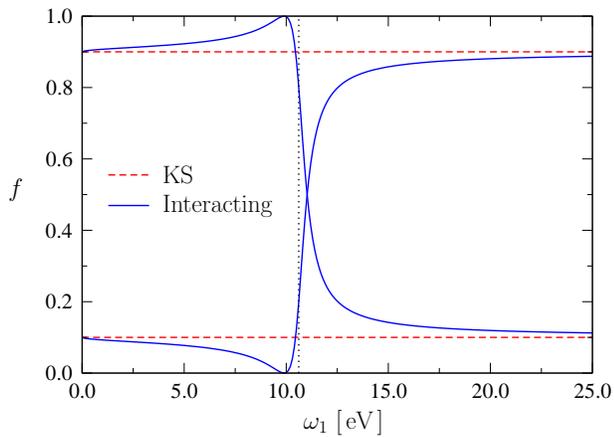}
    }
  \end{picture}
  \caption{Oscillator strengths as function of
  $\omega_1$ .}
   \label{fig3a} 
   \vspace{0.4cm}
\end{figure}
\begin{figure}[t]
  \unitlength1cm
  \begin{picture}(5.5,5.5)
    \put(-3.5,-8.5){
      \includegraphics{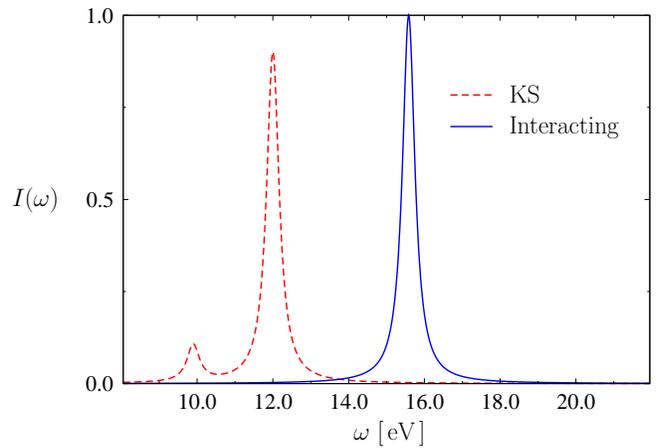}
    }
  \end{picture}
  \vspace{0.2cm}
  \caption{Interacting and Kohn-Sham spectra at the critical value
    $\omega_1 = \omega_d \approx 9.90$ eV.  All intensity is 
    in the upper transition.}
   \label{fig9}
  \vspace{0.4cm} 
\end{figure}

But in Fig. \ref{fig3a}, we plot the associated oscillator
strengths.  The effect of coupling is extremely dramatic.
Note first that, for $\omega_1$ below the strong coupling
region, the bigger peak is {\em enhanced} above its KS
value, and the smaller one reduced.  This is pole repulsion,
and it is felt even very far from the strong coupling region.
This effect is entirely missing from SPA.  Next we see that
there is even a critical value $\omega_d$ ($d$ for {\em dark}) 
at which $f_{-}=0$ exactly.  This means the lower peak 
disappears entirely, and all strength is in the upper peak 
(Fig. \ref{fig9})!  From Eqs. (\ref{physOsc}) and (\ref{thetahi}),
we find
\ben
\Delta\Omega\SPA = g(\alpha\KS)\, |M_{12}|~~~~~(f_1=0),
\een
where $g(\alpha)= 4/\tan2\alpha=16/3$ for $f_1\KS=0.1$
as is the case here.  This yields 8.93 eV, whereas the
exact result is 9.90 eV.  
%Thus this 'magic' separation is also
%of order $M_{12}$ around the SPA crossing point.  

\begin{figure}
  \unitlength1cm
  \begin{picture}(5.5,5.5)
    \put(-3.5,-8.5){
      \includegraphics{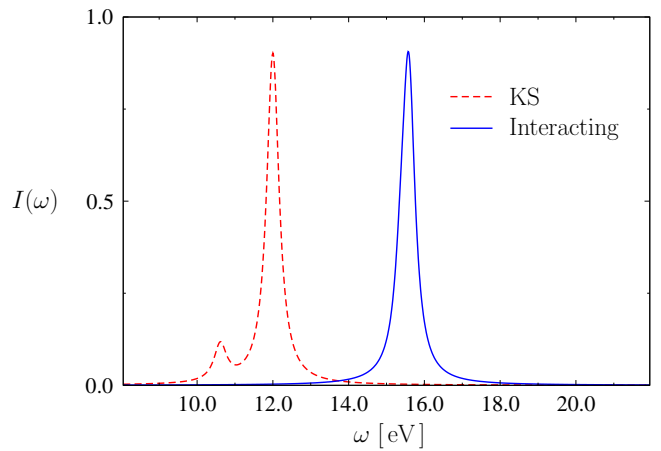}
    }
  \end{picture}
  \vspace{0.2cm}
  \caption{Interacting and Kohn-Sham spectra for $\omega_1 = \omega_c \approx 10.61$ eV.}
  \label{fig6a} 
  \vspace{0.4cm}
\end{figure}

By increasing $\omega_1$ just a little more, we come to the position
of the avoided crossing $\omega_c$ ($c$ for {\em crossing}),
where $\theta=\pi/2$.  In fact, Eqs. (\ref{alphaKS})-(\ref{physOsc}) yield here
\ben
f_\pm = \half \pm \langle f\KS \rangle,
\label{fpi}
\een
where $\langle f\KS\rangle$ denotes the geometric mean, ${\sqrt{f_1\KS\, f_2\KS}}$.
In our case, this  yields $f_{-} = 0.2$ and $f_{+} = 0.8$, respectively, giving 
the lower peak {\em double} its KS weight.
In Fig \ref{fig6a}, we show the spectrum for $\omega_1=\omega_c$, and
observe how much it differs from its KS doppelganger.
There appears to be only one peak, but in fact there are still two,
although the broadening obscures this.  They are very close together.

\begin{figure}[t]
  \unitlength1cm
  \begin{picture}(5.5,5.5)
    \put(-3.5,-8.5){
      \includegraphics{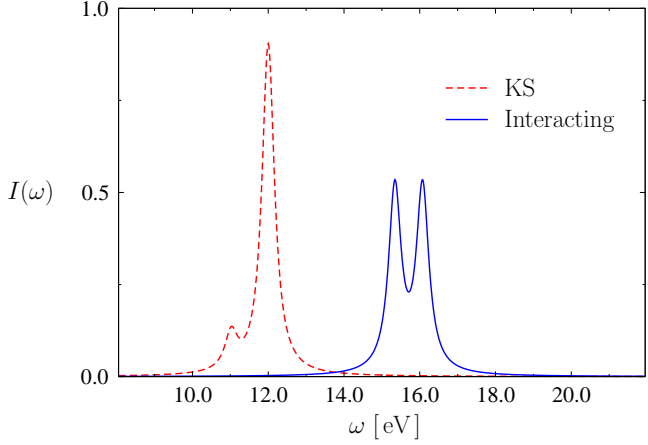}
    }
  \end{picture}
  \vspace{0.2cm}
  \caption{Interacting and Kohn-Sham spectra for $\omega_1 =\omega_e \approx 11.02$~eV, 
  producing equal
  interacting oscillator strengths.}
   \label{fig10} 
   \vspace{0.4cm}
\end{figure}

The final interesting point is $\omega_e$ ($e$ for {\em equal}), where the 
interacting oscillator strengths equal, i.e., both are 1/2. 
At the equality point,
$\alpha=\pi/4$, and so $\theta=\pi/2-2\alpha\KS$.  Again using the
high-frequency limit, Eq. (\ref{thetahi}), yields
\ben
\Delta\Omega\SPA = -4\, M_{12} \cot(2\alpha\KS),
\label{equality}
\een
i.e., the same distance above the crossing point, as the amount the
point $f_-=0$ is below.  This yields 12.29 eV, whereas the exact number
is 11.02 eV (Fig. \ref{fig10}).

\begin{figure}
  \unitlength1cm
  \begin{picture}(5.5,5.5)
    \put(-3.5,-8.5){
      \includegraphics{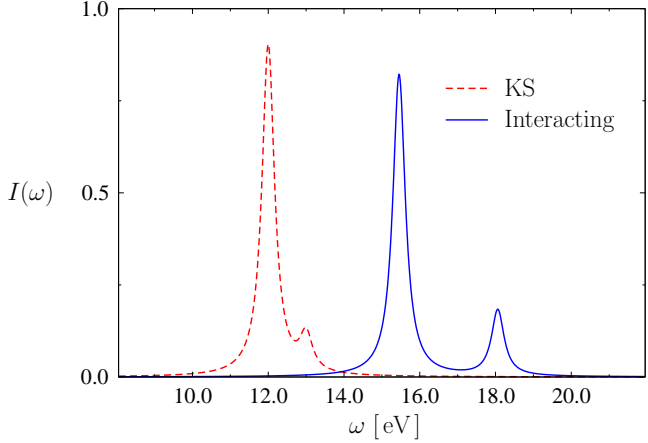}
    }
  \end{picture}
  \vspace{0.2cm}
  \caption{Interacting and Kohn-Sham spectra  for $\omega_1$ = 13~eV.} 
  \label{fig7a} 
\end{figure}

Finally, in Fig. \ref{fig7a}, we consider $\omega_1=13$ eV.  Now the
oscillator strengths have returned (almost) to their KS values, but
$+$ and $-$ have been reversed.
\begin{figure}[ht]
  \unitlength1cm
  \begin{picture}(5.5,5.5)
    \put(-3.5,-8.5){
      \includegraphics{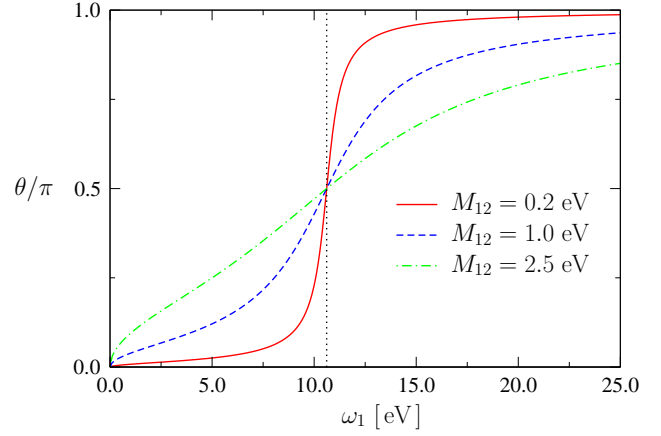}
    }
  \end{picture}
  \vspace{0.2cm}
  \caption{The scaled coupling angle $\theta/\pi$ as function of $\omega_1$. The
  plot compares three different regimes for the off-diagonal matrix element $M_{12}$.}
   \label{fig2b} 
\end{figure}
Lastly, we demonstrate the dependence of these results on the 
strength of $M_{12}$ relative to the diagonal elements.  We have so far
presented only the case $M_{12}  << M_{ii}$.  But we have argued that
it is only the ratio $|M_{12}|/\Delta\Omega\SPA$ that matters.  Thus
increasing $M_{12}$ does not change the shape of the curves (around
the turnover point), but only changes the scale on which the
action takes place.  In Fig. \ref{fig2b}, we change $M_{12}$ to 1 eV and
2.5 eV, and see this occur.  Since the turnover occurs on a scale
of about $4|M_{12}|$, almost the entire region has strong coupling for
$M_{12}$ = 2.5 eV.

\begin{figure}[b]
   \vspace{0.4cm}
  \unitlength1cm
  \begin{picture}(5.5,5.5)
    \put(-3.5,-8.5){
      \includegraphics{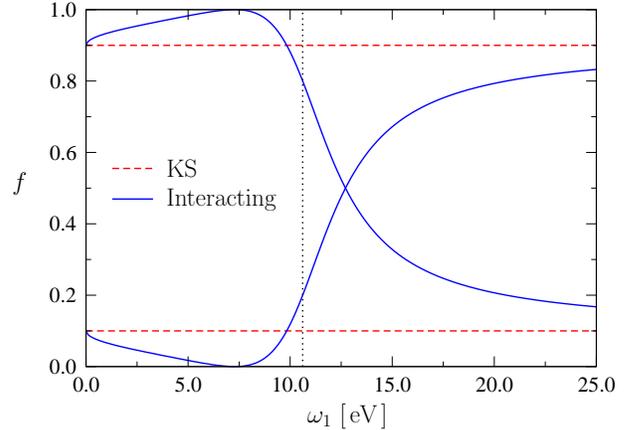}
    }
  \end{picture}
  \vspace{0.2cm}
  \caption{Same as Fig. \ref{fig3a}, but for the case $M_{12} = 1.0$ eV.}
   \label{fig3b} 
\end{figure}

\begin{figure}[ht]
  \unitlength1cm
  \begin{picture}(5.5,5.5)
    \put(-3.5,-8.5){
      \includegraphics{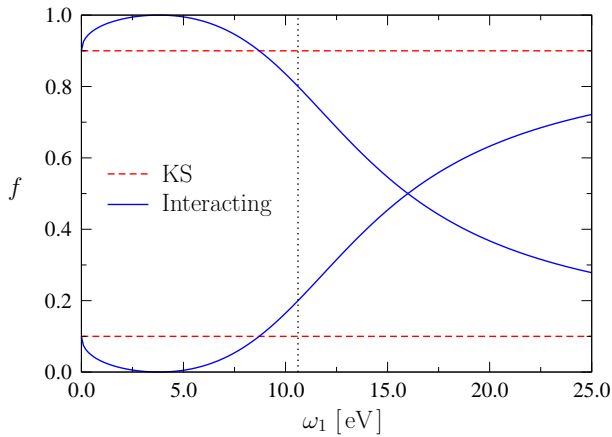}
    }
  \end{picture}
  \vspace{0.2cm}
  \caption{Same as Fig. \ref{fig3a}, but for the case $M_{12} = 2.5$ eV.}
   \label{fig3c} 
   \vspace{0.4cm}
\end{figure}

Lastly, we examine this behavior as a function of $M_{12}$.  In Figs \ref{fig3b}
and \ref{fig3c},  we repeat the plot of oscillator strengths 
versus $\omega_1$ for this system, but
now with $M_{12}=1$ eV and $M_{12}= 2.5$ eV, respectively.  We see that the larger
values lead to qualitatively similar behavior, but over a broader frequency
scale.

\section{Inversion}
\label{s:inverse}

The above sections present the TDDFT response equations in the usual manner.
First solve the ground-state KS problem, finding occupied and
unoccupied levels, then calculate matrix elements of the kernel
(with some functional approximation), and calculate the
true transitions and oscillator strengths of your system.
However, we are motivated to gain insight into the excitations,
and so we ask the reverse question:  Given the experimental spectrum,
what can we learn about the kernel?
Inverting our equations to solve for $\theta$ yields:
\be
\label{thetafinal}
\theta = 2\,(\alpha-\alpha\KS)
\ee
Thus, knowledge of the KS oscillator strengths, combined only with 
the experimental oscillator
strengths, yields the mixing angle, which measures how strongly the transitions
are mixed!
No knowledge of the {\em positions} of transitions
is needed.

Solving for the diagonal matrix elements we arrive at
\bea
\label{Wdiag}
W_{11} & = & \overline{\Omega^2} - (\Delta \Omega^2/2) \cos\theta, \nonumber\\
W_{22} & = &  \overline{\Omega^2} + (\Delta \Omega^2/2) \cos\theta,
\eea
where $\overline{\Omega^2}$ is the average of $\Omega^2$ and 
$\Delta \Omega^2$ is the difference, while the off-diagonal matrix element is
\ben
\label{W12}
W_{12} = (\Delta \Omega^2/2)\; \sin\theta.
\een
Again, the experimental positions combined with the mixing angle
are sufficient to determine the 
elements of the matrix $W$.  The kernel matrix elements themselves are then
found simply, by using the KS transition frequencies:
\ben
\label{Mdiag}
M_{jj} = \frac{W_{jj}}{4\,\omega_j} -\frac{\omega_j}{4}
\een
and 
\be
\label{M12}
M_{12} = \frac{\Delta \Omega^2\; \sin\theta}{8\sqrt{\omega_1\omega_2}}.
\ee
These equations provide an exact way to recover
the matrix elements $W_{ij}$ of the original matrix and therefore the matrix-elements $M_{ij}$ of the 
kernel $K$ solely from the knowledge of the eigenvalues and the angle $\theta$. 

While the above formulas are completely general, in practice strong coupling tends 
to occur between neighboring transitions.  In those cases, the differences
between the two transition frequencies are often much smaller than the transition
frequencies themselves.  Thus we expand in the small parameter
$\Delta\Omega/{\overline\Omega}$, to find
\bea
\label{Wexpansion}
W_{11}&=&{\overline\Omega}\, \left( {\overline\Omega} -\Delta\Omega\; \cos\theta\right),\nonumber\\
W_{22}&=&{\overline\Omega}\, \left( {\overline\Omega} +\Delta\Omega\; \cos\theta\right),\nonumber\\
W_{12}&=&{\overline\Omega}\, \Delta\Omega\; \sin\theta.
\eea
To further extract the matrix elements of the kernel, we assume the
KS transitions satisfy the same requirement, i.e., that the experimental
transitions are close to the KS ones on the scale of the average transition.
This yields:
\bea
\label{Mexpansion}
M_{11}&=&(\overline\Omega-\Delta\Omega\cos\theta)/4-\omega_1/4,\nonumber\\
M_{22}&=&(\overline\Omega+\Delta\Omega\cos\theta)/4-\omega_1/4,\nonumber\\
M_{12}&=&\Delta\Omega/4 \sin\theta.
%M_{11}&=&\Omega_--\omega_1 -\Delta\Omega\; \sin^2\frac{\theta}{2},\nonumber\\
%M_{22}&=&\Omega_+-\omega_2 -\Delta\Omega\; \sin^2\frac{\theta}{2},\nonumber\\
%M_{12}&=& \Delta\Omega\; \sin\theta.
\eea
These simple expressions give the matrix elements directly, once the
KS and experimental information is known.  The mixing angle is determined
completely by the oscillator strengths, as in Eq. (\ref{thetafinal}).
These expressions were used to analyze X-ray absorption spectra in
Ref. \cite{SGAS05}.

\section{Conclusions}
\label{s:conc}

To summarize, we have presented the exact formulas that arise from a double-pole
approximate solution to the TDDFT linear response equations.  We have shown how
these reduce to the single-pole approximation when the coupling between transitions
is weak, and derived the leading terms in this expansion, finding results consistent
with those of Ref. \cite{AGB03}.  However, with DPA, we can go beyond that work, by
considering strong coupling.  We also derive simpler expressions that are valid
when the transitions are of much higher frequency than the splittings.  We illustrated
our results, finding (i) that the oscillator strengths can deviate significantly
from their KS values, even when the coupling is very weak, (ii) that the scale to
compare the off-diagonal matrix element to is the splitting in the single-pole approximation,
and (iii) that the weaker peak even vanishes at a special value of the coupling.

KB acknowledges support of the US Department of Energy under grant number
DE-FG02-01ER45928, and the National Science Foundation under grant
number CHE-0355405. Partial financial support by the EXC!TING Research and
Training Network of the EU, the NANOQUANTA network of excellence and the 
Deutsche Forschungsgemeinschaft within Sfb658 is gratefully acknowledged.

\end{document}